\documentclass[12pt]{iopart}
\usepackage{amsfonts}
\usepackage{graphicx}% Include figure files

\begin{document}

\title{Uncovering the community structure associated with the diffusion dynamics on
networks}

\author{Xue-Qi Cheng and Hua-Wei Shen}

%\author{Hua-Wei Shen and Xue-Qi Cheng\footnote{Author to whom any correspondence should be addressed},}

\address{Institute of Computing Technology, Chinese Academy of Sciences,
Beijing, People's Republic of China}

\eads{\mailto{cxq@ict.ac.cn,shenhuawei@software.ict.ac.cn}}

\begin{abstract}
As two main focuses of the study of complex networks, the community
structure and the dynamics on networks have both attracted much
attention in various scientific fields. However, it is still an open
question how the community structure is associated with the dynamics
on complex networks. In this paper, through investigating the
diffusion process taking place on networks, we demonstrate that the
intrinsic community structure of networks can be revealed by the
stable local equilibrium states of the diffusion process.
Furthermore, we show that such community structure can be directly
identified through the optimization of the conductance of network,
which measures how easily the diffusion occurs among different
communities. Tests on benchmark networks indicate that the
conductance optimization method significantly outperforms the
modularity optimization methods at identifying the community
structure of networks. Applications on real world networks also
demonstrate the effectiveness of the conductance optimization
method. This work provides insights into the multiple topological
scales of complex networks, and the obtained community structure can
naturally reflect the diffusion capability of the underlying
network.
\end{abstract}


\maketitle

As two main focuses of the study of complex networks, the community
structure and the dynamics on networks have both attracted much
attention in various scientific fields. On one hand, many methods
for community detection have been proposed and applied successfully
to some specific complex networks~\cite{Newman2004, Guimera2005,
Newman2006, Palla2005, Rosvall2007, Shen2009a, Blondel2008,
Shen2009b, Fortunato2009}. Each method requires, explicitly or
implicitly, a definition of a community from different perspectives,
such as the centrality measures, link density, percolation theory,
and network compression. Generally, community structure is known as
the existence of groups of nodes such that nodes within a group are
much more connected to each other than to the rest of the
network~\cite{Girvan2002}. Community structure may provide insight
into the relation between the structure and the function of complex
networks. Taking the World Wide Web as an example, closely
hyperlinked web pages form a community and they often concern
related topics~\cite{Kleinberg2001, Cheng2009}.

A well known definition for a community is via the modularity, which
is proposed by Newman \textit{et al} as a quality function for a
partition of the network. Modularity is effective for detecting the
community structure of many real world networks. However, as pointed
out by Fortunato \textit{et al}~\cite{Fortunato2007}, modularity
suffers the resolution limit problem and this problem raises
concerns about the reliability of the communities detected through
the optimization of modularity. In~\cite{Arenas2008b}, the authors
claimed that the resolution limit problem is attributable to the
coexistence of multiple scale descriptions of the topological
structure of the network, while only one scale is obtained through
directly optimizing the modularity. In addition, the definition of
modularity only considers the significance of link density from the
static topological structure of network, and it is unclear how the
modularity based community structure is correlated to the dynamics
on network.

On the other hand, many efforts are devoted to understanding the
properties of the dynamical processes taking place on the underlying
networks~\cite{Boccaletti2006,Arenas2008a}. In recent years,
researchers have begun to investigate the correlation between the
community structure and the dynamics on networks. For example,
Arenas \textit{et al} pointed out that the synchronization reveals
the topological scale in complex networks~\cite{Arenas2006}. In
addition, the random walk on a network was also extensively studied
and used to uncover community structure of the
network~\cite{Rosvall2008, Zhou2004, Pons2005}. In~\cite{Zhou2003a,
Zhou2003b}, the random walk on a network is introduced for defining
the distance between network nodes, and an algorithm based on this
distance is proposed for partitioning the network into communities.
In~\cite{Delvenne2008}, the authors proposed quantifying and ranking
the quality of network partitions in terms of their stability,
defined as the clustered autocovariance in the random walk process
taking place on the network.

In this paper, through investigating the diffusion process taking
place on the network, we note that local equilibrium states appear
before the diffusion process reaches the final equilibrium state.
The stability of local equilibrium states can be measured by their
duration time in the diffusion process. Then, we demonstrate that
the intrinsic community structure is revealed by the stable local
equilibrium states of the diffusion process. Furthermore, we show
that such community structure can be directly identified through the
optimization of the conductance of the network, which measures how
easily the diffusion among different communities occurs.

We first introduce some notations used later. An undirected network
$G=(V,E)$ with $N$ nodes is often described in terms of its
adjacency matrix $A$ whose elements $A_{xy}$ denote the strength of
the link connecting the nodes $x$ and $y$. The strength of node $x$
is denoted by $s_x=\sum_y{A_{xy}}$. For a node set $V_1\subseteq V$,
$|V_1|$ denotes the number of node in $V_1$, the volume of $V_1$ is
defined as $vol(V_1)=\sum_{x\in V_1}s_x$, and
$in\_vol(V_1)=\sum_{x\in V_1,y\in V_1}{A_{xy}}$ is referred to as
the inward volume of $V_1$.

We start with investigating the diffusion process which describes
the dynamics of a random walker moving on the network. At each time
$t$, the random walker moves from its current node $x$ to one of its
neighboring nodes $y$ randomly with the probability $p(x\rightarrow
y)=A_{xy}/s_x$. The dynamics of this process can be described as
\begin{equation}
\frac{d\rho_x(t)}{dt}=-r\sum_{y} L^T_{xy} \rho_y(t) \,, \ \
x=1,\ldots,N\,, \label{eq1}
\end{equation}
where $\rho_x(t)$ is the probability that the random walker resides
at the node $x$ at the time $t$, and $r$ is a parameter controlling
the rate of the diffusion process. The matrix $L$ is the normalized
Laplacian matrix defined as $L=I-D^{-1}A$, where $I$ is the identity
matrix and $D$ is a diagonal matrix with its diagonal elements
$D_{xx}=s_x$.

For any starting node, as time proceeds, the diffusion process
described in the equation~(\ref{eq1}) will definitely move towards
equilibrium if the underlying undirected network is connected and
non-bipartite~\cite{Doh2004}. When the diffusion process is at the
equilibrium state, it satisfies the so-called \emph{detailed balance
condition}~\cite{Barrat2008}, i.e., the probability that the random
walker walks through the nodes $x$ and $y$ successively is equal to
the probability that the random walker walks through the nodes $y$
and $x$ successively. Formally, the detailed balance condition can
be denoted by $\rho_x(t)p(x\rightarrow y)=\rho_y(t)p(y\rightarrow
x)$ and the reduced form is $\rho_x(t)/s_x = \rho_y(t)/s_y$ for
undirected networks.

We explore the transients in the whole diffusion process instead of
only the final equilibrium state. During the diffusion process on a
network, it is known that the detailed balance condition is
satisfied among highly interconnected nodes first and then,
sequentially, among less interconnected ones, until among all the
nodes. In order to evaluate how closely two nodes $x$ and $y$
satisfy the detailed balance condition at the time $t$, we introduce
a measure $c_{xy}(t)$ as
\begin{equation}
c_{xy}(t)=\left\langle
\left|\frac{\rho_x(t)}{s_x}-\frac{\rho_y(t)}{s_y}\right|
\right\rangle,\label{eq2}
\end{equation}
where $\langle\cdots\rangle$ averages over different realizations of
the diffusion process with randomly selected starting nodes. In
practice, a pair of nodes $x$ and $y$ is said to satisfy the
detailed balance condition at the time $t$ when $c_{xy}(t)$ is
smaller than a given threshold. A set $V$ of nodes is said to
satisfy the detailed balance condition if the average value
$\sum_{y\in V}{c_{xy}(t)}/|V|$ of $c_{xy}(t)$ for each node $x$ is
smaller than the given threshold. Relative to the final equilibrium
state, we say that the diffusion process is at a local equilibrium
state when only several groups of nodes locally satisfy the detailed
balance condition. Using the matrix $c_{xy}(t)$, we can trace the
different local equilibrium states during the diffusion process.

\begin{figure}[t]%figure1
\centering
  \begin{tabular}[t]{ccc}
    \hspace{50pt}
    \begin{tabular}[b]{cc}
    \mbox{\includegraphics*[width=.6\textwidth]{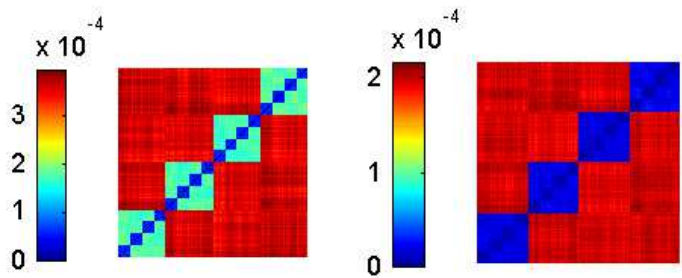}}\\
    \end{tabular}
  \end{tabular}
\vspace{-10pt}
  \caption{(color online). The matrix $c_{xy}(t)$ of two transients in the
  diffusion process taking place on the H13-4 network. Each value of
  $c_{xy}(t)$ is the average over $10,000$ realizations of the diffusion
  process with randomly selected starting nodes. The parameter $r=0.01$. }
  \label{fig1}
\end{figure}

\begin{figure}
\begin{center}
\hspace{50pt}
\includegraphics[width=0.32\textwidth]{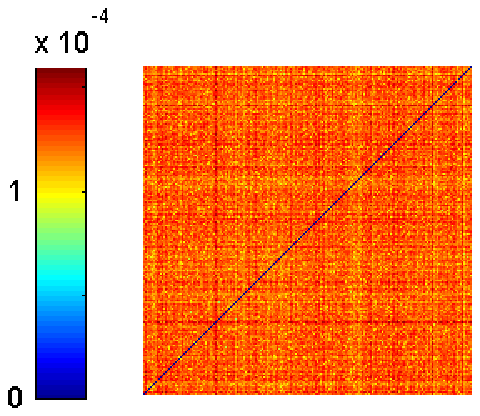}\hspace{5pt}
\includegraphics[width=0.25\textwidth]{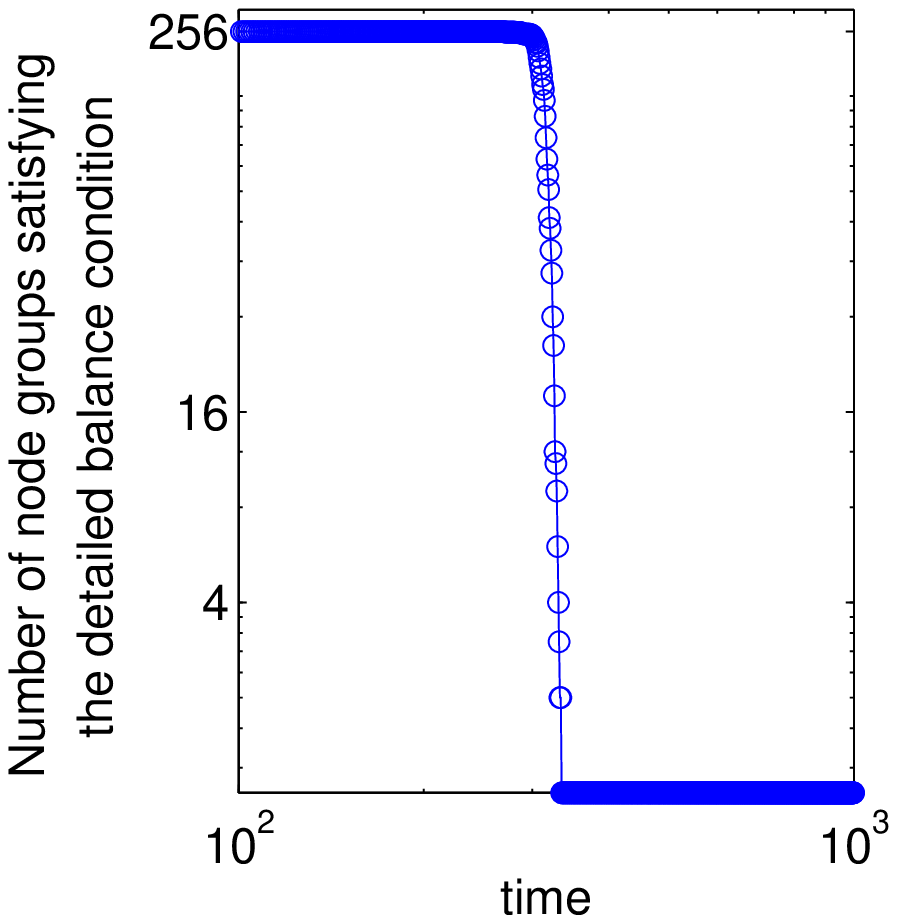}
\\ \hspace{90pt}\mbox{\rule{0pt}{8pt}{\rm (a)}}
\hspace{120pt} \mbox{\rule{0pt}{8pt}{\rm (b)}}
\end{center}\vspace{-10pt}
\caption{(color online). The diffusion dynamics on the randomized
H13-4 network. (a) The matrix $c_{xy}(t)$ of a transient in the
diffusion process. Each value of $c_{xy}(t)$ is the average over
$10,000$ realizations of the diffusion process with randomly
selected starting nodes. The parameter $r=0.01$. (b) The number of
node groups satisfying the detailed balance condition as a function
of time $t$. Here, the threshold for $c_{xy}(t)$ is set to be
$1.0\times 10^{-4}$.} \label{fig2}
\end{figure}

As an example, we use $c_{xy}(t)$ to analyze the diffusion process
on the H13-4 network, which is constructed according
to~\cite{Arenas2006}. The network has two predefined hierarchical
levels, the first hierarchical level consists of $4$ groups of $64$
nodes and the second hierarchical level consists of $16$ groups of
$16$ nodes. Figure~\ref{fig1} illustrates the $c_{xy}(t)$ of two
transients corresponding to two different local equilibrium states
of the diffusion process. The squares along the diagonal suggest
that the corresponding groups of nodes satisfy the detailed balance
condition. These node groups reveal the predefined hierarchical
levels in the H13-4 network. For comparison, we further investigate
the diffusion dynamics on the randomized H13-4 network, which is
constructed through shuffling the edges of the H13-4 network used in
figure~\ref{fig1}. From figure~\ref{fig2}(a), we can see that there
is no node group locally satisfying the detailed balance.

A phenomenon similar to that illustrated in figure~\ref{fig1} has
also been observed in the synchronization process.
In~\cite{Arenas2006}, the authors claimed that this phenomenon
reveals the topological scale of networks. The authors also pointed
out that the phenomenon of the synchronization process is correlated
with the spectrum of the Laplacian matrix associated with the
underlying network. According to the characteristics of the
Laplacian matrix, as pointed out in~\cite{Luxburg2007}, the
community structure revealed by the synchronization process is
heavily affected by the heterogeneous degree distribution and the
community size distribution. In the following, we will show that the
local equilibrium state phenomenon is correlated with the spectrum
of the normalized Laplacian matrix, which takes the heterogeneous
degree and community size distribution into account and thus behaves
better than the Laplacian matrix as regards clustering the nodes of
network~\cite{Luxburg2007}.

In this paper, a local equilibrium state is regarded as stable if
the set of node groups satisfying the detailed balance condition
remains unchanged for a long duration in the diffusion process. To
investigate the stability of the local equilibrium states, we study
the solution of equation (\ref{eq1}) in terms of the normal modes
$\varphi_i(t)$, which reads
\begin{equation}
\rho_x(t)=\sum_{i}{U_{xi}\varphi_i(t)}=\sum_{i}{U_{xi}\varphi_i(0)e^{-\lambda_irt}}
\,, \ \ x=1,\ldots,N\,,\label{eq3}
\end{equation}
where the $\lambda_i$ are the eigenvalues of the transpose of the
normalized Laplacian matrix $L$, and $U$ is the eigenvector matrix
whose $i$th column is the eigenvector $u_i$ corresponding to the
eigenvalue $\lambda_i$. Given the starting node of the diffusion
process, the initial amplitudes $\varphi_i(0)$ can be determined
according to equation~(\ref{eq3}) due to the eigenvector matrix $U$
being fixed, i.e., $\varphi_i(0)$ only depends on the starting node
of the diffusion process. Note that, as pointed out following the
equation~(\ref{eq2}), we investigate the average behavior of many
different diffusion processes with randomly selected starting nodes.
Thus, the choice of the starting nodes does not affect the results
of the following analysis in this paper. Without loss of generality,
we rank these eigenvalues in the ascending order
$0=\lambda_1\leq$$\lambda_2\leq$$\cdots\leq$$\lambda_i\leq$$\dots$
$\leq\lambda_N$.

As time proceeds in the diffusion process, these normal modes
$\varphi_i(t)=\varphi_i(0)e^{-\lambda_irt}(i\neq 1)$ will decay to
zero. We use $\tau_i$ to denote the time when the normal mode
$\varphi_i(t)$ decays to zero. Formally, $\tau_i$ is infinite. In
practice, a threshold $\varepsilon$ is usually used to determine
when $\varphi_i(t)$ decays to zero, i.e.,
$\varphi_i(t)<\varepsilon$. In this case, we have
\begin{equation}
\tau_i = \frac{1}{\lambda_i}\times
\frac{ln\varphi_i(0)-ln\varepsilon}{r}.\label{eq4}
\end{equation}
All these moments $\tau_i$ ($1\leq i\leq N$) form a series of time
intervals, respectively $[\tau_{N+1}=0,\tau_N)$, $[\tau_N,
\tau_{N-1})$, $\cdots$, $[\tau_{i+1}, \tau_{i})$, $\cdots$,
$[\tau_3, \tau_2)$, $[\tau_2, \tau_1=\infty)$. These time intervals
divide the whole diffusion process into $N$ stages. Specifically,
the time interval $[\tau_{i+1}, \tau_{i})$ is regarded as the $i$th
stage. When the diffusion process is at the $i$th stage, only the
normal modes $\varphi_j(t)$($1\leq j\leq i$) have not decayed to
zero. Thus we have
\begin{equation}
\rho_x(t)\approx\sum_{j=1}^{i}{U_{xj}\varphi_j(t)} \,, \ \
x=1,\ldots,N\,.\label{eq5}
\end{equation}
This indicates that the value $\rho_x(t)$ of node $x$ at the $i$th
stage can be represented by the $i$-dimension coefficient vector of
$\varphi_j(t)$, i.e., $(U_{x1},U_{x2},\cdots,U_{xj}, \cdots,
U_{xi})$. According to equations (\ref{eq2}) and (\ref{eq5}), given
a threshold, we can identify the node groups satisfying the detailed
balance condition through clustering the normalized $i$-dimension
vectors of $(U_{x1},U_{x2},\cdots,U_{xj}, \cdots, U_{xi})$ using,
for example, the $k$-means clustering method. The set of such node
groups is unchanged due to the non-decayed normal modes being fixed
during the same stage. This means that the local equilibrium state
is stable if the corresponding stage persists for a long time.

\begin{figure}[t]%figure1
\centering
 % \item[]
  \begin{tabular}[t]{ccc}
    \vspace{5pt}
    \begin{tabular}[b]{cc}
    \hspace{60pt}
    \mbox{\includegraphics*[width=.55\textwidth]{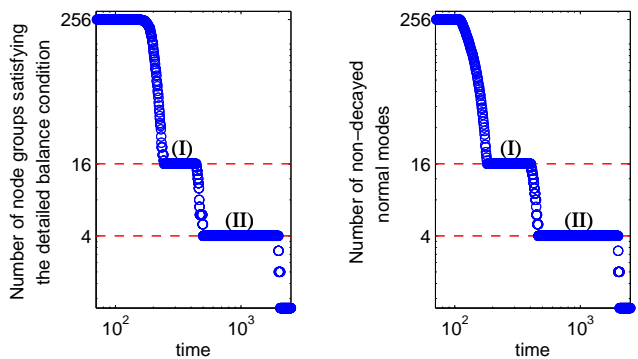}}\\
    \end{tabular}
  \end{tabular}
 % \end{indented}
  \caption{(color online). Left: the number of node groups satisfying the
  detailed balance condition as a function of time $t$. Here, the threshold
  for $c_{xy}(t)$ is set to be $1.0\times 10^{-4}$. Right: the number
  of non-decayed normal modes in terms of the time $t$.}
  \label{fig3}
\end{figure}

Taking the H13-4 network as an example, the left panel of
figure~\ref{fig3} shows the different local equilibrium states of
the diffusion process and the right panel illustrates the different
stages in terms of the number of non-decayed normal modes of the
diffusion process. Through comparing the two panels, we see that
each time one normal mode decays to zero, the diffusion process
changes from a local equilibrium state to a new one. It is observed
that two stable local equilibrium states with long durations emerge
in the diffusion process. The node groups corresponding to these two
states clearly reveal the intrinsic community structure of the H13-4
network. In addition, from figure~\ref{fig2}(b), we can see that no
stable local equilibrium state appears in the diffusion process
taking place on the randomized H13-4 network. This is reasonable, as
it is commonly believed that randomized network have no community
structure. All these findings suggest that the appearance of stable
local equilibrium states in a diffusion process indicates the
existence of community structure in the underlying network.

Note that the diffusion occurs much more frequently within the node
groups than among them when the diffusion process is at a stable
local equilibrium state. This indicates that there exists a high
transitive cohesion inside such node groups. The community structure
comprised of these node groups could well reflect the diffusion
dynamics on the underlying network. Regarding such community
structure as a partition of a network, we measure the quality of the
partition through introducing the \emph{conductance} of networks,
which reflects how easily the diffusion occurs among different
communities.

For a given partition $\mathcal {P}=\{V_1,\ldots,V_k\}$, the
conductance is defined as the average \emph{departure probability},
$p_{dept}(V_i)$, of all the communities $V_i$, that is
$C(\mathcal{P})=\frac{1}{k}\sum_{i=1}^{k}{p_{dept}(V_i)}$. The
departure probability of a community $V_i$ is the probability that
the random walker departs from $V_i$ in the next time step given
that it resides at $V_i$ when the diffusion process is at the final
equilibrium state. Formally, the departure probability of $V_i$ can
be computed by using
\begin{equation}
p_{dept}(V_i)=\frac{\sum_{x\in V_i,y\in
\overline{V}_i}{\rho_x(\infty)p(x\rightarrow y)}}{\sum_{x\in
V_i}{\rho_x(\infty)}}=1-\frac{in\_vol(V_i)}{vol(V_i)},\label{eq6}
\end{equation}
where $\rho_x(\infty)=s_x/vol(V)$ is the stationary distribution
which characterizes the final equilibrium state of the diffusion
process. In this way, the conductance is formally denoted by
\begin{equation}
C(\mathcal
{P})=\frac{1}{k}\sum_{i=1}^{k}{\left(1-\frac{in\_vol(V_i)}{vol(V_i)}\right)}.\label{eq7}
\end{equation}

Actually, it can be proved that the community structure associated
with the stable local equilibrium state can be exactly identified
through minimizing the conductance directly. Without loss of
generality, we assume that the stable local equilibrium state
emerges at the $k$th stage in the diffusion process. As mentioned
above, the community structure associated with this state can be
identified through clustering the normalized $k$-dimension vectors
of $(U_{x1},U_{x2},\cdots,U_{xi}, \cdots, U_{xk})$. Further, through
the matrix trace maximization method~\cite{Yu2003}, it can be proved
that the optimization of the conductance for a fixed $k$ can be done
through clustering the top $k$ eigenvectors of the transpose of the
normalized Laplacian matrix, corresponding to the $1$st to $k$th
columns of $U$. Therefore, the optimization of conductance provides
an effective way to identify the community structure associated with
the stable local equilibrium state.

Now we clarify the difference between the conductance used in this
paper and the earlier measure for the quality of network partition
from the perspective of a random walk on networks. Firstly, as
pointed out in~\cite{Evans2009}, as a measure of the quality of
network partition, the modularity can be described as the difference
between the probability that a random walker resides in the same
community on two successive time steps and the probability that two
independent random walkers both resides in the same community.
Secondly, in~\cite{Delvenne2008}, through considering the random
path with length $t$ instead of the length $1$ for the modularity
and the length of infinity for the spectral partition, the authors
proposed that the stability of network partitions be defined as the
clustered autocovariance of the random walk. This stability provides
a general framework for quantifying and ranking the quality of
network partitions from the perspective of the whole network, i.e.,
it characterizes the fraction of the within-community random paths
with length $t$ with respect to all the random paths of length $t$.
However, the conductance considers the quality of network partition
from the perspective of each community instead of the whole network,
i.e., it reflects the fraction of within-community random paths with
respect to all the random paths departing solely from the community
considered. This provides the advantage for handling the
heterogeneous distribution of community size (or volume) which is
common to real world networks. As follows, the application of the
conductance optimization method to the benchmarks of Lancichinetti
et al also demonstrates that our method can effectively handle the
heterogeneous distribution of community size.

To test the effectiveness of our method for community detection
based on the optimization of conductance, we utilize the benchmark
proposed by Lancichinetti et al in~\cite{Lancichinetti2008}. This
benchmark provides networks with heterogeneous distributions of node
degree and community size. Thus it poses a much more severe test of
community detection algorithms than standard benchmarks. Many
parameters are used to control the generated networks in this
benchmark: the number of nodes $N$, the average node degree $\langle
k\rangle$, the maximum node degree max$\rule[-1pt]{0.15cm}{0.3pt}k$,
the mixing ratio $\mu$, the exponent of the power law node degree
distribution $t_1$, the exponent of the power law distribution of
community size $t_2$, the minimum community size
min$\rule[-1pt]{0.15cm}{0.3pt}c$, and the maximum community size
max$\rule[-1pt]{0.15cm}{0.3pt}c$. In our tests, we use the default
parameter configuration where $N=1000$, $\langle k\rangle=15$,
max$\rule[-1pt]{0.15cm}{0.3pt}k=50$, $t_1=2$, $t_2=1$,
min$\rule[-1pt]{0.15cm}{0.3pt}c=20$, and
max$\rule[-1pt]{0.15cm}{0.3pt}c=50$. By tuning the parameter $\mu$,
we test the effectiveness of our method on the networks with
different fuzziness of communities. The larger the parameter $\mu$,
the fuzzier the community structure of the generated network. In
addition, we adopt the normalized mutual information
(NMI)~\cite{Danon2005} in order to compare the partition found by
the algorithms with the answer partition. The larger the NMI, the
more effective the tested algorithm.

\begin{figure}
\begin{center}
\hspace{60pt}\includegraphics[width=0.28\textwidth]{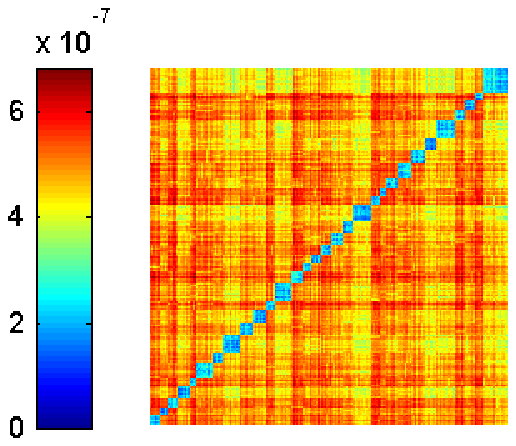}
\includegraphics[width=0.2\textwidth]{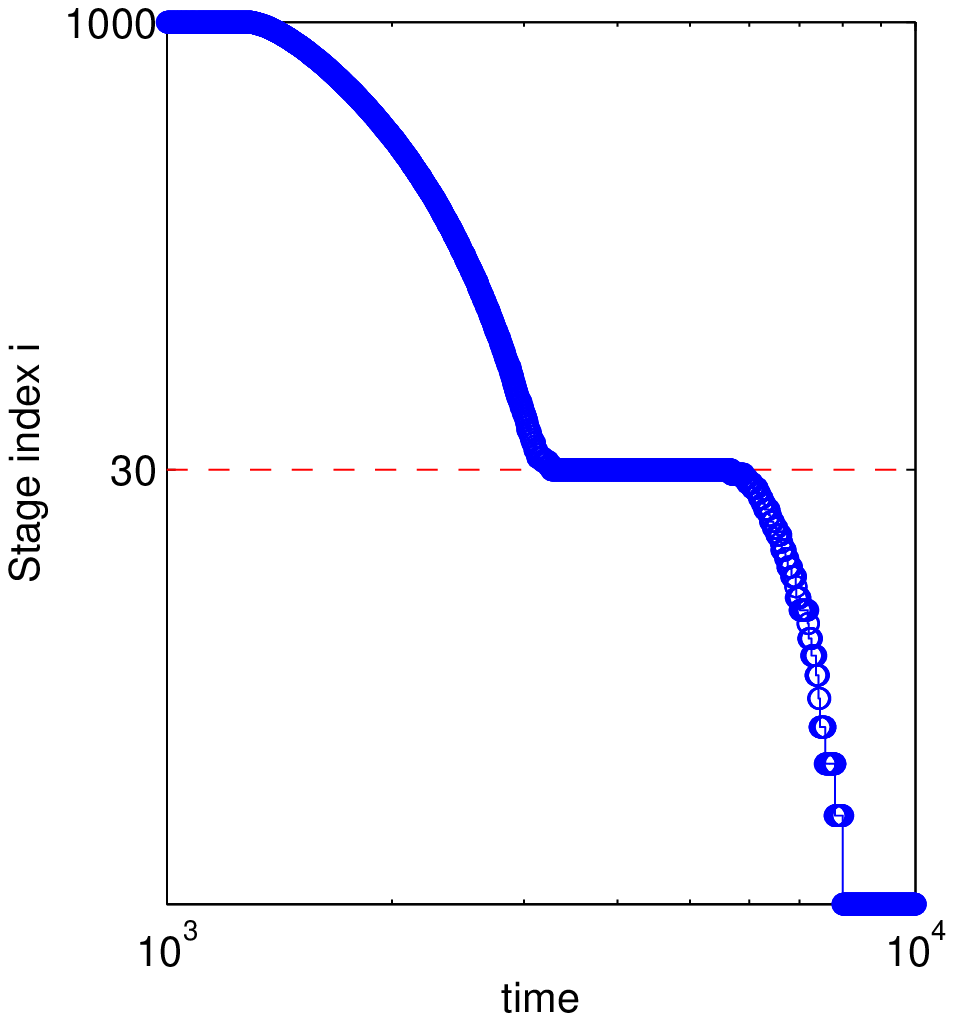}\hspace{15pt}
\includegraphics[width=.3\textwidth]{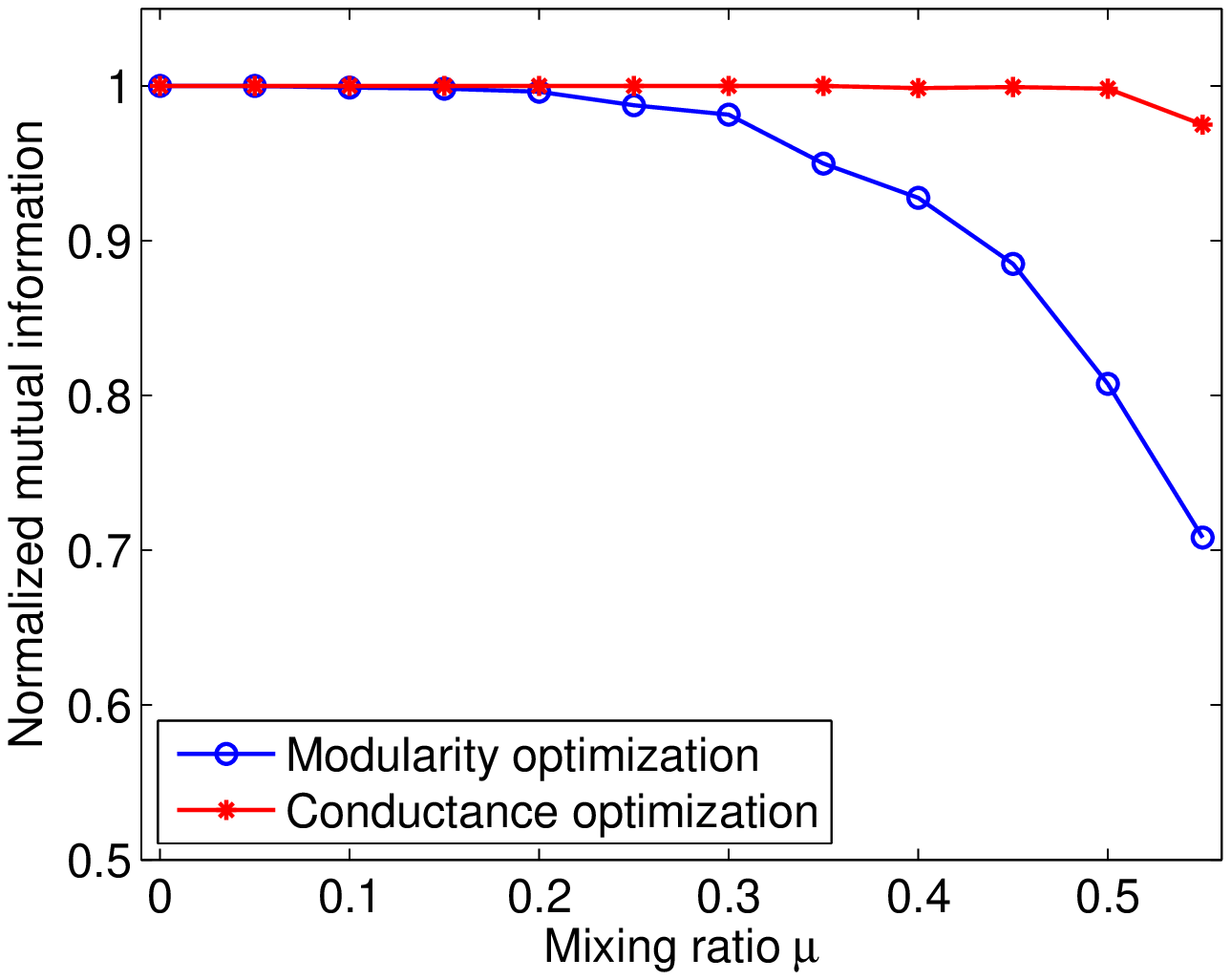}
\\ \hspace{60pt}\mbox{\rule{0pt}{8pt}{\rm (a)}}
\hspace{100pt} \mbox{\rule{0pt}{8pt}{\rm (b)}} \hspace{100pt}
\mbox{\rule{0pt}{8pt}{\rm (c)}}
\end{center}\vspace{-10pt}
\caption{(color online). (a-b) The most stable local equilibrium and
the different stages of the diffusion process on the benchmark
network with the mixing ratio $\mu$=$0.3$ and the number of
communities is $30$. (c)Comparison between the conductance
optimization method and the modularity optimization method on
benchmark networks. Each point corresponds to an average over $100$
network realizations.} \label{fig4}
\end{figure}

Figure~\ref{fig4}(a-b) illustrate the most stable local equilibrium
state and the different stages of the diffusion process on the
benchmark network with the mixing ratio $\mu$=$0.3$ and the number
of communities equal to $30$. The squares along the diagonal
indicate the predefined communities in the network. The number of
these communities is clearly revealed by the most stable local
equilibrium state. Figure~\ref{fig4}(c) shows the comparison between
the conductance optimization method and the modularity optimization
method in terms of the NMI on the benchmark network. When the
community structure is evident, both our method and the modularity
optimization method (e.g., the fast unfolding
algorithm~\cite{Blondel2008} and the spectral
method~\cite{Newman2006}) can accurately identify the community
structure. However, when the community structure becomes fuzzier,
the performance of the modularity optimization method deteriorates
while our method still achieves rather good results.

\begin{figure}
\begin{center}
\hspace{60pt}\includegraphics[width=0.48\textwidth]{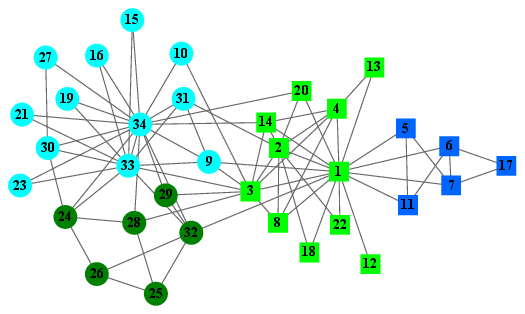}\hspace{30pt}
\includegraphics[width=0.28\textwidth]{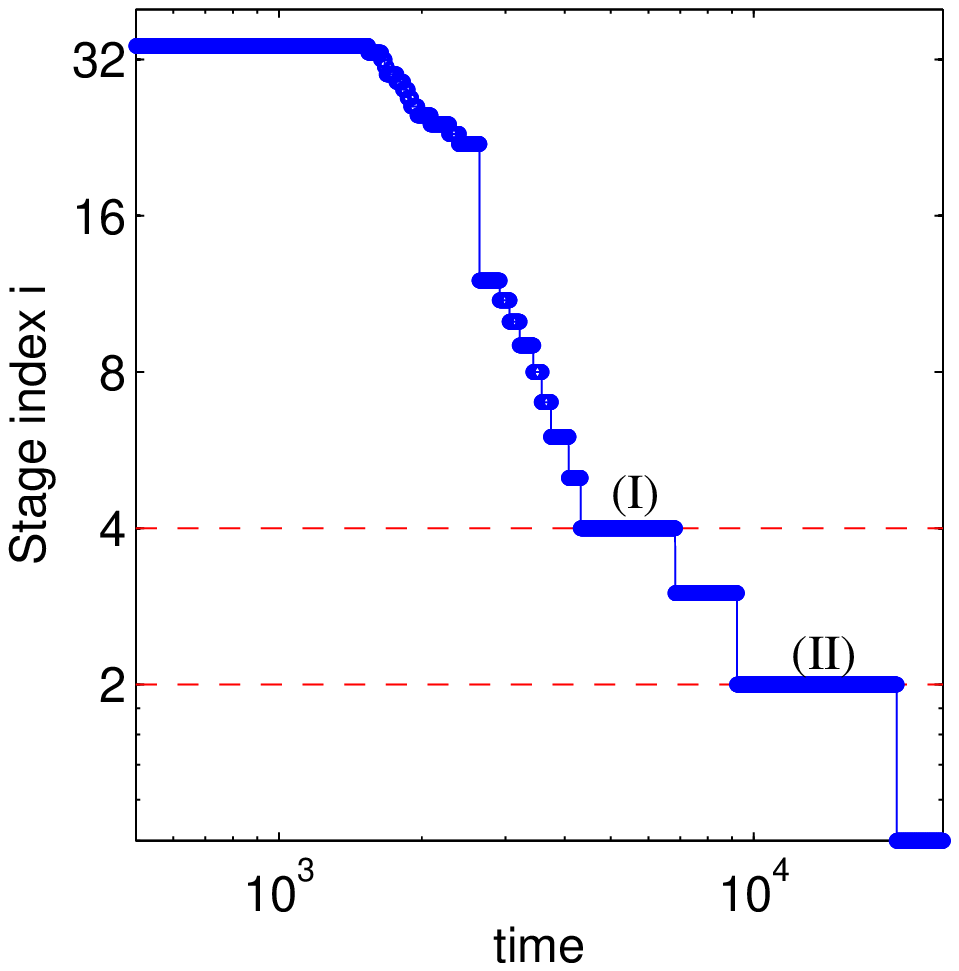}
\\ \hspace{100pt}\mbox{\rule{0pt}{8pt}{\rm (a)}}
\hspace{190pt} \mbox{\rule{0pt}{8pt}{\rm (b)}}
\end{center}\vspace{-10pt}
\caption{(color online). Illustration of the conductance
optimization method using the real world network. (a) The friendship
network of the karate club studied by Zachary~\cite{Newman2004}.
Colors are used to differentiate the communities uncovered by the
conductance optimization method when considering the stage (I) in
figure~\ref{fig5}(b). Shapes, circle and square, are used to
distinguish the communities corresponding to the stage (II) and the
real social split of this network observed by Zachary. (b) The
different stages of the diffusion process taking place on the
network in figure~\ref{fig5}(a). Two most stable equilibrium states
are marked with (I) and (II).} \label{fig5}
\end{figure}

In addition, we also tested the conductance optimization method on
many real world networks which are widely used to evaluate community
detection methods. These networks include the social network of
Zachary's karate club, the social network of dolphins of Lusseau
\textit{et al}, the college football network of the United
States~\cite{Girvan2002}, the journal index network constructed
in~\cite{Rosvall2007}, and the network of political
books~\cite{Newman2006}. For all these networks, the conductance
optimization method obtains extremely good results. Taking Zachary's
network as an example, figure~\ref{fig5}(b) illustrates the
different stages of the diffusion process taking place on the
network. The two most stable local equilibrium states are marked (I)
and (II), and the corresponding communities are depicted in
figure~\ref{fig5}(a). Besides the stages (I) and (II), another
relatively stable state can be also observed during the diffusion
process, as shown in figure~\ref{fig5}(b). The corresponding three
communities are respectively the one comprised of all the circle
nodes and two communities formed by the square nodes but with
different colors, as shown in figure~\ref{fig5}(a). Actually, as
regards the three stable states, it is really hard to say which the
best one is. In this paper, the duration time of each state may
provide an effective candidate measure for the significance of
network divisions.

In summary, we find that several stable local equilibrium states
emerge during the diffusion process on networks with community
structure. These stable states reveal the intrinsic community
structure of the underlying networks. We further propose a
conductance optimization method to identify the community structure,
which naturally reflects the diffusion capability of the network.
This work provides new insights into the number of communities and
the multiple topological scales of complex network. As future work,
we will study the relation and difference between the conductance
and the modularity method.

\ack{This work was funded by the National Natural Science Foundation
of China under grant number $60873245$ and $60933005$. The authors
gratefully acknowledge S Fortunato for useful suggestions. The
authors thank M-B Hu, R Jiang, G Yan and T Zhou for helpful
discussions. The authors also thank the anonymous reviewers for
valuable comments to this paper.}


\section*{References}
\begin{thebibliography}{30}

\bibitem{Newman2004}
M.~E.~J. Newman and M.~Girvan, Phys. Rev. E \textbf{69},
026113 (2004).

\bibitem{Guimera2005}
R.~Guimer\`a and L.~A.~N. Amaral, Nature(London) \textbf{433}, 895
(2005).

\bibitem{Newman2006}
M.~E.~J. Newman, Proc. Natl. Acad. Sci. U.S.A. \textbf{103},
8577 (2006).

\bibitem{Palla2005}
G.~Palla, I.~Der\'enyi, I.~Farkas, and T.~Vicsek, Nature(London)
\textbf{435}, 814 (2005).

\bibitem{Rosvall2007}
M.~Rosvall and C.~T.~Bergstrom, Proc. Natl. Acad. Sci. U.S.A.
\textbf{104}, 7327 (2007).

\bibitem{Shen2009a}
H.~W.~Shen, X.~Q.~Cheng, K.~Cai and M.~B.~Hu, Physica A
\textbf{388}, 1706 (2009).

\bibitem{Blondel2008}
V.~D.~Blondel, J.-L.~Guillaume, R.~Lambiotte and E.~Lefebvre, J.
Stat. Mech.: Theory and Exp., (2008) P10008.

\bibitem{Shen2009b}
H.~W.~Shen, X.~Q.~Cheng, and J.~F.~Guo, J. Stat. Mech.: Theory and
Exp., (2009) P07042.

\bibitem{Fortunato2009}
S.~Fortunato, Phys. Rep. \textbf{486}, 75-174 (2010).

\bibitem{Girvan2002}
M.~Girvan and M.~E.~J. Newman, Proc. Natl. Acad. Sci. U.S.A.
\textbf{99}, 7821 (2002).

\bibitem{Kleinberg2001}
J.~Kleinberg and S.~Lawrence, Science \textbf{294}, 1849 (2001).

\bibitem{Cheng2009}
X.~Q.~Cheng, F.~X.~Ren, S.~Zhou and M.~B.~Hu, New J. Phys.
\textbf{11}, 033019 (2009).

\bibitem{Fortunato2007}
S.~Fortunato and M.~Barth\'{e}lemy, Proc. Natl. Acad. Sci. U.S.A.
\textbf{104}, 36 (2007).

\bibitem{Arenas2008b}
A.~Arenas, A.~Fern\'{a}ndez and S.~G\'{o}mez, New. J. Phys.
\textbf{10}, 053039 (2008).

\bibitem{Boccaletti2006}
S.~Boccaletti, V.~Latora, Y.~Moreno, M.~Chavez, and D.~-U.~Hwang,
Phys. Rep. \textbf{424}, 175 (2006).

\bibitem{Arenas2008a}
A.~Arenas, A.~D\`{i}az-Guilera, J.~Kurths, Y.~Moreno, and C.~Zhou,
Phys. Rep. \textbf{469}, 93 (2008).

\bibitem{Arenas2006}
A.~Arenas, A.~D\`{i}az-Guilera, C.~J.~Perez-Vicente, Phys.
Rev. Lett. \textbf{96}, 114102, (2006).

\bibitem{Rosvall2008}
M.~Rosvall and C.~T.~Bergstrom, Proc. Natl. Acad. Sci. U.S.A.
\textbf{105}, 1118 (2008).

\bibitem{Zhou2004}
H.~Zhou and R. Lipowsky, \textit{Lecture Notes in Computer Science}
\textbf{3038}, 1062 (2004).

\bibitem{Pons2005}
P.~Pons and M. Latapy, \textit{Lecture Notes in Computer Science} \textbf{3733},
284 (2005).

\bibitem{Zhou2003a}
H.~Zhou, Phys. Rev. E \textbf{67}, 041908 (2003).

\bibitem{Zhou2003b}
H.~Zhou, Phys. Rev. E \textbf{67}, 061901 (2003).

\bibitem{Delvenne2008}
J.-C. Delvene, S.~N.~Yaliraki, and M.~Barahona, arXiv:0812.1811.


\bibitem{Doh2004}
J.~D.~Noh and H.~Rieger, Phys. Rev. Lett. \textbf{92}, 118701, (2004).

\bibitem{Barrat2008}
A.~Barrat, M.~Barth\'{e}lemy, and A.~Vespignani, \textit{Dynamical
Processes on Complex Networks} (Cambridge University Press,
Cambridge, UK) (2008).

\bibitem{Luxburg2007}
U. von Luxburg, Stat. Comput. \textbf{17}, 395 (2008).

\bibitem{Yu2003}
S.~X.~Yu and J.~Shi, in Proceedings of the Ninth IEEE International
Conference on Computer Vision (IEEE Computer Society, Washington DC, 2003), pp. 313--319.

\bibitem{Evans2009}
T.~S.~Evans and R. Lambiotte, Phys. Rev. E \textbf{80}, 016105
(2009).

\bibitem{Lancichinetti2008}
A.~Lancichinetti, S. Fortunato, and F. Radicchi, Phys. Rev.
E \textbf{78}, 046110 (2008).

\bibitem{Danon2005}
L.~Danon, J.~Duch, A.~D\`{i}az-Guilera, J.~Duch and A.~Arenas, J.
Stat. Mech., P09008 (2005).

\end{thebibliography}
\end{document}